\documentclass[preprint]{aastex}

\slugcomment{}

\shorttitle{New Constraints on the Short-Term Variability of WR 46}
\shortauthors{V. H\'{e}nault-Brunet et al.}

\begin{document}


\title{New Constraints on the Origin of the Short-Term Cyclical\\Variability of the Wolf-Rayet Star WR~46}


\author{V. H\'{e}nault-Brunet\altaffilmark{1}, N. St-Louis\altaffilmark{2}, S.V. Marchenko\altaffilmark{3}, A.M.T. Pollock \altaffilmark{4}, S. Carpano\altaffilmark{5}, A. Talavera\altaffilmark{4}}

\email{vhb@roe.ac.uk, stlouis@astro.umontreal.ca, sergey.marchenko@ssaihq.com, andy.pollock@esa.int, scarpano@rssd.esa.int, antonio.talavera@sciops.esa.int}


\altaffiltext{1}{Scottish Universities Physics Alliance (SUPA), Institute for Astronomy, University of Edinburgh, Royal Observatory Edinburgh, Blackford Hill, Edinburgh EH9 3HJ, UK.}

\altaffiltext{2}{D\'epartement de Physique, Universit\'e de Montr\'eal, and Centre de Recherche en Astrophysique du Qu\'ebec (CRAQ), C.P.~6128, Succ.~Centre-Ville, Montr\'eal, Qu\'ebec H3C 3J7, Canada.}

\altaffiltext{3}{Science Systems and Applications, Inc., 10210 Greenbelt Road, Suite 600, Lanham, MD 20706, USA.}

\altaffiltext{4}{European Space Agency, XMM-Newton Science Operations Centre, ESAC, 28691 Madrid, Spain.} 

\altaffiltext{5}{Research and Scientific Support Department, ESTEC/ESA, PO Box 299, 2200, AG Noordwijk, the Netherlands.}


\begin{abstract}

The Wolf-Rayet star WR~46 is known to exhibit a very complex variability pattern on relatively short time scales of a few hours. Periodic but intermittent radial velocity shifts of optical lines as well as multiple photometric periods have been found in the past. Non-radial pulsations, rapid rotational modulation or the presence of a putative low-mass companion have been proposed to explain the short-term behaviour. In an effort to unveil its true nature, we observed WR~46 with {\it FUSE} (Far Ultraviolet Spectroscopic Explorer) over several short-term variability cycles. We found significant variations on a time scale of $\sim$8~hours in the far-ultraviolet (FUV) continuum, in the blue edge of the absorption trough of the \ion{O}{6}~$\lambda\lambda$1032,~1038 doublet P~Cygni profile and in the \ion{S}{6}~$\lambda\lambda$933,~944 P~Cygni absorption profile. We complemented these observations with X-ray and UV light-curves and an X-ray spectrum from archival {\it XMM-Newton} (X-ray Multi-Mirror Mission - Newton Space Telescope) data. The X-ray and UV light-curves show variations on a time scale similar to the variability found in the FUV. We discuss our results in the context of the different scenarios suggested to explain the short-term variability of this object and reiterate that non-radial pulsations is the most likely to occur.

\end{abstract}



\section{Introduction}

\noindent
The Wolf-Rayet (WR) star WR~46  (HD 104994, DI Cru) is well-known for its remarkable short-term variability. All the suspected causes of this variability (a close, low-mass binary companion, rapid rotational modulation, non-radial pulsations) represent very rarely observed phenomena for Wolf-Rayet stars. The confirmation of either of these scenarios would therefore provide insight into our understanding of the physics and evolution of massive stars.

Despite the fact that some scenarios have been favoured to explain the puzzling behaviour of this star, we still lack a satisfying explanation of the underlying mechanism. Because of the detection of radial velocity variations, the presence of an unseen close binary companion was initially a popular suggestion. Although it runs into many difficulties, this possibility has never been completely excluded up to now. It has also been suggested that multi-frequency non-radial pulsations of a single WR star could cause the observed wind variability \citep{veenc, oliveira2004}. While this last scenario appears to be at least part of the solution, more observations in selected spectral windows were needed to progress towards a clearer picture of this enigmatic star.

It is for this reason that we obtained a time-resolved series of {\it FUSE (Far Ultraviolet Spectroscopic Explorer)} observations and that we retrieved from the archive available {\it XMM-Newton} (X-ray Multi-Mirror Mission - Newton Space Telescope) data of WR~46. {\it FUSE} observed in the far-ultraviolet (FUV) spectral range, giving access to P~Cygni profiles of resonant transitions from various species. These profiles are characteristic of dense stellar outflows and the study of their variability enables us to probe the evolution of the wind structure of WR~46 as a function of time. We were particularly interested in the high ionization potential \ion{O}{6}~$\lambda\lambda$1032,~1038 doublet, which, along with the X-ray data from {\it XMM-Newton}, could provide information on energetic phenomena such as accretion onto a compact companion or wind-embedded shocks.

Before we present these observations, we first highlight in \S\ref{hist} key aspects of the rich but complex observational history of WR~46 in order to place our own data set in the general context. In \S\ref{obs_sec} we describe our observations and the data reduction procedures. In \S\ref{analysis}, we present our analysis of the data, and in \S\ref{discuss} we discuss our results. Our conclusions can be found in \S\ref{conc}.

\section{Observational history \label{hist}}

\subsection{Physical properties \label{prop}}

WR~46 is classified as WN3p in the Seventh Catalogue of Galactic WR Stars \citep{vanderhucht01}. It is characterized by the presence of strong \ion{N}{5} and \ion{He}{2} lines and the absence of hydrogen. The ``p" stands for peculiar and denotes the presence of unusually strong \ion{O}{6}~$\lambda \lambda$3811,~3834 emission lines \citep[e.g.][]{crowther95}, whose occurence is probably due to the susceptibility of that doublet to low density \citep{veen02b}. Although it has once been considered a low mass X-ray binary \citep{niemela95} or a V~Sagittae star \citep{steinerdiaz98}, \citet{crowther95}, \citet{marchenko2000}, \citet{veen02b,veenc}, \citet{oliveira2004} and \citet{gosset} gave convincing arguments in favour of the Wolf-Rayet nature of WR~46. As a weak-lined WN early (WNE) star \citep{crowther95}, it exhibits a low degree of atmospheric extension ($R_{*}\sim R_{\tau_{R}=2/3}$) and triangular emission-line profiles indicating very low wind densities. Weak-lined WNE stars like WR~46 are as hot as the strong-lined WNE stars, but with mass-loss rates about an order of magnitude lower.

WR~46 is a known X-ray source whose discovery with {\it Einstein} was reported by \citet{pollock87}. Later, based on {\it ROSAT} PSPC observations \citep{wessolowski95}, \citet{crowther95} estimated the X-ray luminosity of WR~46 to be $L_{\rm X}=5\pm2\times10^{31}$~erg~s$^{-1}$ for a distance of 4.0~kpc \citep{tovmassian} and noted that this is typical of single WN stars \citep{pollock95}. In their detailed analysis of {\it XMM-Newton} data of WR~46, \citet{gosset} deduced a luminosity of $L_{\rm X}$(0.2$-$10.0 keV)$=7.7\times10^{32}$~erg~s$^{-1}$ (for $d=$4.0~kpc). Using our own X-ray model (see \S\ref{XMM_spec}) shows the previous {\it Einstein} and {\it ROSAT} X-ray measurements to be consistent with the X-ray luminosity from {\it XMM} data within about 10\%. \citet{gosset} interpreted the X-ray spectrum of WR~46 as dominated by a soft component but also detected a harder component above 3~keV, similar to the X-ray spectrum of other presumably single WN stars \citep{skinner2002a, skinner2002, ignace2003}.

In the present work, we adopt the results from a recent CMFGEN \citep{hillier98} blanketed wind model of WR~46 (Crowther, private communication). The stellar and wind parameters, estimated from a combination of nitrogen and oxygen optical/UV line diagnostics, are presented in Table \ref{param}. Note that for the wind terminal velocity, we adopt the value of 2775~km~s$^{-1}$ measured from the saturated \ion{O}{6}~$\lambda\lambda$1032,~1038 P Cygni absorption line \citep{Willis2004}, a resonance doublet that in principle allows to trace material further out in the wind (at lower densities, closer to the actual terminal velocity) than the non-resonance optical/UV lines used in other studies. The adopted mass was derived using the mass-luminosity relation of \citet{schaerer92}.

\subsection{Variability}

Since \citet{monderen88} first reported the photometric variability of WR~46, many studies have presented time-resolved photometric and/or spectroscopic follow-ups of this star. We refer to some of the more recent studies \citep[e.g.][]{veena, oliveira2004, gosset} for a detailed review of the variability observed in WR~46. Here, we just outline the main periods found by various authors and the conclusions that were drawn about the origin of the variability.  Figure \ref{periods} shows the many photometric and spectroscopic (radial-velocity) periods reported, along with the time scales determined in this work (see \S\ref{fuv_lc} and \S\ref{xmmlc}). The spectroscopic ``period'' from this work is not a radial-velocity period, but it is instead the time scale of variations found in the FUV \ion{O}{6}~$\lambda\lambda$1032,~1038 doublet P~Cygni profile. The different periods and time scales observed suggest that the period of WR~46 is changing and/or that multiple periods are present. However, we stress that, ideally, the reality of each of the secondary (i.e. non-dominant) periods shown on Figure \ref{periods} would need to be confirmed with a long-term and very intense data set.

Although the first observations of short-term variability in WR~46 (binary-like double-wave light-curve, low-amplitude radial-velocity variations with $K\sim$100~km~s$^{-1}$) led to the suggestion that it was a binary with a low-mass companion \citep[e.g][]{vangenderen1991, veen95}, subsequent observations really casted doubt on this hypothesis. The fact that the dominant period appears to change over time \citep{veena} or that the radial-velocity variations unexpectedly disappear in about one out of three consecutive nights \citep{marchenko2000, veen02b} is difficult to reconcile with a binary scenario.
While not completely discarding the possibility of a close binary, the most recent studies of WR~46 \citep{veenc, oliveira2004} proposed that non-radial pulsations may well play a dominant role in the short-term variability of this star.

In addition to its short-term cyclical variability, WR~46 is known to exhibit significant variability on a time scale of months. A brightness increase occurred between 1989 and 1991, followed by a decrease from 1991 to 1993 \citep{marchenko98, veena}. From a series of {\it IUE} spectra, \citet{veen02b} inferred that the 1991 brightness maximum was accompanied by an increase of the mass-loss rate. \citet{oliveira2004} found evidence for a difference in the degree of ionization of WR~46 between June 1996 and January 2002 and suggested that this is probably caused by a variation of the global density of the wind. The link between the possible evolution of the short-term period(s) and the long-term brightness and global wind density variations however remains unclear.

\section{Observations and data reduction \label{obs_sec}}

\subsection{{\it FUSE}}

The {\it FUSE} scientific instrument has been described in detail by \citet{moos2000}. The instrument imaged four spectra (SiC1, SiC2, LiF1, LiF2) split in two segments (A and B), for a total of eight segments covering the spectral window from 905 to 1187 \AA\ with substantial overlap between individual segments.

We secured {\it FUSE} observations of WR~46 under guest observer time (program E113, PI: N. St-Louis) in order to carry-out time-resolved FUV spectroscopy over many cycles of variability of the star. In total, for the three main consecutive observations of this program in March 2006 (E1130103, E1130104, E1130105), the star was followed for more than 30 hours spread over a timespan of about 50 hours. A previous observation of WR~46 was presented by \citet{Willis2004} in their {\it FUSE} atlas of WR stars, and a few other shorter observations of this star are also found in the {\it FUSE} archive. We retrieved all available science observations of WR~46 for the analysis presented in this work. The log of the {\it FUSE} observations is presented in Table \ref{log}, which gives the {\it FUSE} data set number, the date and time of the beginning and end of the observation, the aperture and mode used, as well as the number of exposures and the total exposure time of each observation. Note that the observations were obtained in time-tag (TTAG) mode, for which the arrival times of photons are recorded. 

Each individual raw exposure was run through the latest version of the {\it FUSE} calibration pipeline, CalFUSE v3.2 \citep{dixon07}. The pipeline first corrects the input data for ``gaing-sag'', ``event bursts'', and positional shifts of detected photons due to various effects. Then, a correction for background light is applied and a one-dimensional spectrum is extracted. Finally,  the eight channel spectra are calibrated in wavelength and flux. In order to maximize the time coverage, we did not mask the data obtained during the daytime portion of the orbit as is often done for {\it FUSE} data. Geocoronal airglow emission lines are stronger for day-time data, but this did not represent a serious constraint here. The strongest airglow line, \ion{H}{1} Lyman-$\beta$ at 1026 \AA, is located right in the saturated part of the absorption component of the \ion{O}{6} $\lambda\lambda$1032, 1038 P Cygni profile and therefore it does not mask any precious spectral information. The other airglow lines seen in our data were much weaker.

Since we wanted photometric exposures to later extract accurate light-curves, we modified CalFUSE output intermediate data files (IDF) to remove the time intervals where the count rate in the spectral extraction window drops relatively abruptly due to the drifting of the star out of the aperture or away from the aperture center. It is well known that the photometric accuracy of {\it FUSE} is influenced by various effects that cannot be fully corrected by the CalFUSE pipeline \citep{dixon07}. A target centered in an aperture of the guide channel (LiF 1 prior to July 2005, LiF 2 after that) can be misaligned in the apertures of the other three channels. Also, with the loss of reaction wheels (the first two in 2001), channel drifts can even temporarily move the target out of the guide-channel aperture, as happened for some exposures of WR~46. The pipeline attempts to flag times when the target is out of the aperture, but this time lost to pointing excursions is by default underestimated to avoid the rejection of good data. A careful examination and screening of the IDF allowed us to insure the photometric quality of the data. New spectra were then extracted from these IDF for each exposure separately. The individual exposure times of these spectra are typically on the order of a thousand seconds. We attempted to further break down all the exposures in small pieces of a few hundred seconds, but came to the conclusion that the marginal improvement in time resolution was generally not worth the associated decrease in the signal-to-noise ratio of spectra.

The spectra were extracted from the same modified IDF that were used to obtain light-curves. Because the count rate during the masked intervals is lower than normal and because these masked intervals usually represent a minor fraction of the exposure time (especially for the guide channel apertures), the fraction of spectral photons lost due to the screening of the IDF is small. Thus, the signal-to-noise ratio of the extracted spectra is not significantly affected. In a few cases, a major fraction if not all of the exposure had to be masked and no spectrum was extracted, but then the original spectrum that was extracted previous to screening for pointing losses had a very low signal-to-noise ratio and was hardly usable anyway. Using photometric data from the modified IDF to extract spectra advantageously gives a more reliable absolute flux calibration, thus making it possible to study long-term changes in the FUV spectrum of WR~46.

When possible, an instrumental artifact nicknamed the ``worm" was removed from the data. This feature is caused by electron repeller grid wires that prevent light from the target from reaching the detector, resulting in a wide artificial absorption trough. It was most noticeable in the LiF1B spectral segment of the observations we retrieved. It was also sometimes seen in the LiF2A segment. To remove the LiF1B (LiF2A) worm of a given exposure while keeping intact the smaller scale spectral features of the LiF1B (LiF2A) spectrum, we multiplied the LiF1B (LiF2A) spectrum by the smoothed ratio of the overlapping LiF2A (LiF1B) and LiF1B (LiF2A) spectra. This method was not always possible, as for some exposures a worm was apparent in both LiF2A and LiF1B. In these cases, the LiF1B and LiF2A spectral segments of these exposures were not considered in our analysis of variability.

To first isolate spectral line profile variability from continuum variability, all spectra were scaled to a common continuum level. The spectra were also co-aligned to the spectrum of a reference exposure by cross-correlating them over small spectral ranges that include narrow interstellar absorption lines. The alignment required for the different spectra was typically of the order of 0.04 \AA\ or less. All the spectral segments from the reference exposure were previously co-aligned on the guide-channel (LiF 2) spectral segments, for which the wavelength calibration errors are minimal \citep{dixon07}. For each exposure, we then performed a weighted average merging of the flux data in all eight channels, with the weights inversely proportional to the square of the statistical errors. In the end, we obtained a single 1-D, heliocentric corrected, flux-calibrated spectrum for each exposure (except when the exposure suffered from serious misalignment of the target).

Light-curves were obtained to investigate the FUV continuum variability. Other light-curves were also extracted to look at variable spectral-line regions (see \S\ref{tvs_sec}) integrated over broader wavelength intervals than the typical resolution of {\it FUSE} spectra, but with a much better time-resolution than it would be possible by just looking at our time-series of spectra. The routine ttag\_lightcurve (see Bernard Godard's IDF cookbook\footnote{http://archive.stsci.edu/fuse/analysis/idfcook/}) was used to extract light-curves from the modified IDF discussed above. We estimated 1-$\sigma$ uncertainties on the data points of the light-curves by assuming Poisson statistics ($\sigma=\sqrt{N}$, where $N$ is the number of counts).

Specifically, three different types of light-curves were produced. The first shows the variations in the FUV continuum. The spectral window used for these FUV continuum light-curves includes all wavelength intervals that do not encompass the emission lines and P Cygni profiles identified in \S\ref{fuse_spec}, the strongest predicted airglow emission lines \citep[see][]{feldman}, the edges of the wavelength range of each spectral segment, where systematic errors in the flux are significant \citep{sahnow2000}, and narrow interstellar absorption features. The two other types of light-curves show the variations in the \ion{O}{6} $\lambda\lambda$1032, 1038 and \ion{S}{6} $\lambda\lambda$933, 944 doublet P Cygni absorption components, the only two variable spectral regions in the {\it FUSE} spectrum of WR~46 (see \S\ref{tvs_sec}.). The \ion{O}{6} light-curve was extracted using the wavelength region from 1019.5 to 1022.5 \AA, and the \ion{S}{6} light-curve the wavelength regions from 924.5 to 929 \AA\ and 934 to 939 \AA. These light-curves are slightly affected by continuum variability, but they are largely dominated by the line variability, which is much stronger than the continuum contribution in these selected spectral regions. This was easily checked, for example by realizing that the variability in these spectral regions was roughly the same for the spectra that were scaled to a common continuum level compared to the unscaled series of spectra.

For each extracted light-curve, the contributions from all possible spectral segments were combined. For the FUV continuum light-curve, we combined the continuum light-curves of all eight {\it FUSE} spectral segments since continuum variations appeared to be the same across the entire {\it FUSE} spectral range. For the \ion{O}{6} $\lambda\lambda$1032, 1038 P Cygni absorption light-curve, all the spectral segments covering the spectral range concerned (namely LiF1A, LiF2B, SiC1A, and SiC2B) were used. Similarly, spectral segments SiC1B and SiC2A, covering the absorption component of the \ion{S}{6} $\lambda\lambda$933, 944 P Cygni profile, were used for the third light-curve. Note that when a worm feature was present in a spectrum, the light-curve of the affected spectral segment and exposure was not used.

To combine the light-curves of individual spectral segments we first normalized their average count rate to 1 for each observation and rebinned them into time intervals of 0.2 hours, propagating Poisson 1-$\sigma$ errors accordingly. An average light-curve of all the individual spectral segments concerned, weighted using errors, was then computed. To double-check the photometric quality of the data, we made sure that the light-curves from individual channels appeared correlated before averaging them.

\subsection{{\it XMM-Newton} \label{xmmdat}}

WR~46 was observed by {\it XMM-Newton} \citep{jansen2001} between 02:05:52 and 23:22:57 on 2002 February 8 (ObsID 0397\_0109110101; 2XMM catalogue source name = 2XMM J120518.7-620310) with the EPIC and OM instruments. As discussed in \S\ref{prop}, the {\it XMM-Newton} X-ray spectrum of WR~46 is presented by \citet{gosset}, along with X-ray light-curves in different bands extracted from the {\it XMM-Newton} data. To inform our discussion of the short-term variability of WR~46, we also retrieved the publicly available {\it XMM-Newton} data of this star, and performed the additional task of extracting UV light-curves simultaneous to the X-ray light-curve.

The EPIC observations presented here were obtained with the pn detector used in the extended full frame mode \citep{struder} and with the two MOS detectors used in the full frame mode \citep{turner}. A medium filter to reject parasite optical/UV light from the target was used for all three detectors. The OM instrument was used in the imaging mode with the broadband ultraviolet filters UVM2 and UVW2 (effective wavelength 231 and 212 nm respectively).

Custom X-ray source and background spectra were extracted from the calibrated event lists with standard procedures in the XMM-Newton Science Analysis System (SAS) v9.0.0, using circular source and annular background selection regions. The light-curve was calculated with a PSF modeling procedure to estimate simultaneously source and background variations, whose results agreed well with the more approximate standard methods. This light-curve includes the whole {\it XMM-Newton} energy band. Although we decided to calculate our own spectra and light-curves, they proved to be consistent with those available through the XMM-Newton Science Archive via the 2XMM catalogue.

The OM observations consisted of 25 exposures with the UVM2 filter and 30 
exposures with the UVW2 in image mode with 1000 s exposure time for all of
them except the last 10 exposures with UVW2 that had only 860~s$\times$5 and
800~s$\times$5. Aperture photometry of the target was performed using the SAS. 
The source was automatically detected by the SAS in all
images and the count rate was measured using an aperture of radius 35 pixels
(circa 17.5 arc sec). The background was measured in an annulus centered on the 
target. The measured rates were corrected from coincidence losses and time
sensitivity degradation of the detector by the SAS.

In order to later analyze the full OM UV light-curve, we put one filter into the system of the other using the procedure described below. To perform this correction, we have to assume that the total flux ratio in the band pass of the UVM2 and UVW2 filters is constant throughout the observations. This assumption seems reasonable given that the color variations in the $VBLUW$ photometric monitoring of \citep{veena} are smaller than 1\%, and the amplitude of photometric variability is around $\pm$5\% in both the UV (see \S\ref{xmmlc}) and the optical \citep[e.g.][]{veena}. We then have:

\begin{equation}
\frac{F_{\rm UVW2}}{F_{\rm UVM2}} = K,
\label{}
\end{equation}

\noindent{where $F$ refers to the flux and $K$ is a constant. From the calibration of the OM, we know that:

\begin{equation}
F_{\rm filter} = R_{\rm filter} \ C_{\rm filter},
\label{}
\end{equation}

\noindent{where $F$ is the flux, $R$ is the count rate and $C$ is a rate to flux conversion factor. This conversion factor can be obtained from the effective area and observations of standard stars, from the OM calibration documentation, or from the SAS Current Calibration File (CCF).

To compute $K$, we used the series of observations of WR~46 in the UVW2 and UVM2 filters as follows:

\begin{equation}
K  = \frac{\bar{R}_{\rm UVW2} \ C_{\rm UVW2}}{\bar{R}_{\rm UVM2} \ C_{\rm UVM2}},
\label{}
\end{equation}

\noindent where $\bar{R}_{\rm UVW2}$ and $\bar{R}_{\rm UVM2}$ are the mean count rates in each filter. The normalization to convert the measured rate in the UVM2 filter to the UVW2 equivalent is then

\begin{equation}
R_{\rm UVW2} = K \ R_{\rm UVM2} \frac{C_{\rm UVM2}}{C_{\rm UVW2}}.
\label{}
\end{equation}

\noindent Here, we have $\bar{R}_{\rm UVW2}=152.417$ counts~s$^{-1}$, $\bar{R}_{\rm UVM2}=292.748$~ counts~s$^{-1}$, $C_{\rm UVW2}=5.75\times10^{-15}$~erg/cm$^2$/\AA/count, and $C_{\rm UVM2}=2.20\times10^{-15}$~erg/cm$^2$/\AA/count. The same method was applied to compute normalized errors.

\section{Analysis \label{analysis}}

\subsection{Time-averaged {\it FUSE} spectrum \label{fuse_spec}}

From a visual inspection of the time-averaged {\it FUSE} spectrum of WR~46 (see Figure \ref{tvs}, top panels) and the synthetic interstellar transmission spectra for the {\it FUSE} range computed, for example, by \citet{Willis2004}, it is clear that the spectrum of WR~46 is highly contaminated by ISM features. The column density of interstellar H$_{2}$ is perhaps as high as 10$^{18}$ atoms/cm$^{2}$ and maybe even more. The interstellar H$_{2}$ and H-Lyman lines effectively remove a large fraction of the stellar flux shortward of 1020 \AA. Despite the plethora of interstellar lines in the {\it FUSE} spectrum of this star, we focus in this work on the FUV stellar wind features. Note that we did not attempt a detailed modeling of the {\it FUSE} spectrum. We did consider a model of the FUV spectrum of WR~46 (Crowther, private communication), but the direct comparison with the {\it FUSE} spectrum proved to be difficult given the wealth of interstellar absorption lines, so we did not try to fine-tune this model. It is however worth mentioning that the slope of the FUV continuum in the reddened model (E(B-V)=0.34, see \citealt{crowther95}) was in rough agreement with the {\it FUSE} spectrum.

WR~46 (WN3p) is the earliest WN subtype covered in the {\it FUSE} atlas of WR stars of \citet{Willis2004}. The substantial interstellar contamination makes it very difficult to identify the expected \ion{S}{6} $\lambda\lambda$933.4, 944.5, \ion{C}{3} $\lambda$977, and \ion{N}{3} $\lambda$991 resonance lines. \citet{Willis2004} noted that the latter two are clearly not strong in emission or as P~Cygni profiles, but that the \ion{S}{6} doublet seems to be present as a P~Cygni profile. We can indeed identify the absorption troughs of this P Cygni profile around $\sim$925 \AA\ and $\sim$935 \AA. Note that this \ion{S}{6} resonance doublet is also observed as a P Cygni profile in the {\it FUSE} spectrum of the other WN3 star (HD 32109, WN3b) of the atlas of \citet{Willis2004}.  HD 32109 is in the LMC and suffers far less ISM contamination, which makes the identification of this spectral feature much easier.

\citet{Willis2004} also reported that the \ion{He}{2}~$\lambda 1085$, \ion{P}{5} $\lambda\lambda$1118, 1128, and possibly \ion{Fe}{6} $\lambda 1168$ lines appear present as weak emissions or P~Cygni profiles in WR~46. We also note the presence of a relatively wide absorption trough around 1060 \AA\ which might be the absorption component of a currently unidentified weak P~Cygni profile. The {\it FUSE} spectrum of WR~46 is dominated by a very strong saturated P~Cygni profile in the \ion{O}{6} $\lambda\lambda$1032, 1038 resonance doublet. Given the high level of ionization potential of species in the wind, \citet{Willis2004} attributed the formation of \ion{O}{6} in this case to normal photoionization as opposed to Auger-ionization formation in shocked wind gas. More generally, they considered the  ``super ions" of \ion{O}{6} and \ion{S}{6} as photoionized wind features for WN3-WN6 stars, while they are probably the result of Auger ionization in WN7-WN9 stars, and probably absent in WN10-WN11 stars.

The presence of species of high ionization potential (138 eV for \ion{O}{6}) in the {\it FUSE} spectrum of WR~46 is not surprising given the very high effective temperature of this star. It is also in agreement with what is observed in its {\it IUE} spectrum \citep{willis1986} which is dominated by \ion{N}{5} $\lambda\lambda$1238, 1242 and \ion{O}{5} $\lambda$1371 P Cygni profiles and \ion{He}{2} $\lambda 1640$ emission, but in which no significant \ion{N}{3} emission is detected. {\it IUE} observations of WR~46 were also presented by \citet{crowther95} who noted that the \ion{He}{2} $\lambda$1640, \ion{N}{4} $\lambda$1718, and \ion{C}{4} $\lambda\lambda$1548, 1550 lines are observed as pure emission features, with \ion{N}{4} and \ion{C}{4} extremely weak, while lines from higher excitation stages such as \ion{N}{5} $\lambda\lambda$1238, 1242 and \ion{O}{5} $\lambda1371$ are observed as P~Cygni profiles. The optical spectrum of WR~46 is also dominated by emission lines from the high-ionization species \ion{N}{5} and \ion{O}{6}, and also from \ion{He}{2} \citep[e.g.][]{oliveira2004}.

\subsection{Time-series analysis of {\it FUSE} spectra \label{tvs_sec}}

To statistically determine which parts of the {\it FUSE} spectrum of WR~46 show significant spectral line variability, we performed a temporal variance spectrum analysis \citep[TVS;][]{fullerton96} on the spectra from all the different observations summarized in Table~\ref{log}. We did this separately for each {\it FUSE} channel, which allowed us, by comparing the results, to distinguish more easily real variable features from spurious variations.  It was also easier to assess the quality of individual spectra by proceeding channel by channel. The co-addition of several {\it FUSE} spectral segments (with two to four segments overlapping depending on the wavelength region) introduces more variations of the continuum noise across the spectrum, which would complicate the determination of the weights in the calculation of the TVS (see below). Prior to computing the TVS of each channel, we rebinned the spectra onto a common linear wavelength scale with a 0.13 \AA\ bin size. We refer to \citet{fullerton96} for the details of how to compute the TVS. We simply note here that the continuum noise was estimated by fitting a linear function to selected regions of the continuum for each channel, and that the errors on the flux of each pixel were taken from the flux error estimates output by CalFUSE, obviously modified according to the rebinning of the data.

We present in Figure \ref{tvs} (bottom panels) the TVS$^{1/2}$ spectrum for each channel. The quantity TVS$^{1/2}$ scales linearly with the size of the spectral flux deviations and is therefore more practical to use than the TVS itself because it gives a direct estimate of the amplitude of the spectral variations. Note that when plotting the TVS, we masked the wavelength intervals where peaks were caused by variability in airglow emission lines. We also masked spurious peaks caused by isolated outlying data points. These were common in the wings of narrow interstellar lines, since  the TVS is very sensitive to small residual wavelength misalignment in the presence of steep spectral gradients.

The dashed horizontal line plotted with the TVS$^{1/2}$ spectrum of each channel indicates the 99\% statistical significance value for that channel. We can see from Figure \ref{tvs} that there is significant variability in the blue edge of the \ion{O}{6} $\lambda\lambda$1032, 1038 P Cygni absorption components and also, to a lower level, in the \ion{S}{6} $\lambda\lambda$933, 944 P Cygni absorption components. These lines are in fact the two strongest P Cygni profiles present in the {\it FUSE} spectrum of WR~46. To show the evolution of these variable features with time for all {\it FUSE} observations of WR~46, we present in Figure \ref{gray} a gray-scale plot of the differences between each individual spectrum (with the spectral segments from all channels combined) and the reference time-averaged spectrum for the wavelength regions where significant variability was detected. In this figure, the bottom panels display each difference profile as an intensity profile. The top panel presents the time-averaged spectrum (thick line) with the spectra from all exposures overplotted (thin lines). For each line, we also overlay a velocity scale to show the velocity of the variable feature with respect to the center of the line.

We can see that the variability in each component of the \ion{S}{6} doublet P Cygni absorption occurs mainly for velocities ranging roughly from -0.6 $v_{\infty}$ to a little faster than -$v_{\infty}$, while the variability in the saturated \ion{O}{6}  P Cygni absorption occurs in excess of -$v_{\infty}$, up to approximately -1.25 $v_{\infty}$. From the gray-scale plot of the consecutive observations of March 2006, we also see that the spectral features evolve on time scales of several hours, although the variability pattern is clearly not strictly periodic. It is indeed difficult to conclude anything about this variability pattern just by looking at the sequence of spectra. We do not clearly see displacement of spectral features in velocity space as a function of time, especially for the \ion{S}{6} doublet. The variations in each component of the \ion{S}{6} doublet absorption seem more or less correlated with each other. When the variable features are different in the two components, it is possibly due to the fact that the velocity intervals severely affected by interstellar absorption are also different for the two components. Finally, by comparing the evolution of the \ion{O}{6} and \ion{S}{6} profiles, we see no obvious correlation between the variability patterns in these features.

Recall that for the above analysis, all the spectra were scaled to a common continuum level to focus on spectral-line variability. We can see from this analysis that the spectral lines do not seem to vary in strength over the years spanned by our {\it FUSE} observations. To identify possible long-term variability in the continuum, we looked at series of spectra that were not scaled to a common continuum level and found no evidence for significant long-term variations in the FUV continuum.

\subsection{{\it FUSE} light-curves and period search \label{fuv_lc}}

The {\it FUSE} light-curves extracted with the method described in \S\ref{obs_sec} are presented in Figure \ref{FUV_lca} and Figure \ref{FUV_lcb}. Times are given in hours from the beginning of each observation (from the beginning of observation E1130103 for the March 2006 observations). As previously mentioned, these light-curves were extracted to provide a measure of the FUV continuum variability and of the integrated variability in selected spectral regions. Since the spectral variations generally appeared as broad features in the gray-scale plot of Figure \ref{gray}, we believe that the light-curves for which the count rate across the whole variable spectral region is integrated represent a good measure of the global line variability.

As we can see from Figures \ref{FUV_lca} and \ref{FUV_lcb}, the amplitude of the FUV continuum light-curve is typically around $\pm$5\%. The amplitude of the \ion{O}{6} P Cygni absorption blue edge light-curve is typically around $\pm$20\%. The situation of the \ion{S}{6} P Cygni absorption light-curve is similar to that of the \ion{O}{6} light-curve, although the velocity intervals over which the changes occur are different. Interestingly, the amplitude of the variability in the P Cygni absorption components for both the \ion{O}{6} and \ion{S}{6} lines seems larger when the amplitude of the FUV continuum variability is also larger. We note some similarity in the shape of the three light-curves, although they do not appear strictly correlated. We will explore possible correlations between the different light-curves in \S\ref{corr}.

To search for periodicity in the variability pattern of WR~46, we applied the fast Lomb-Scargle algorithm of \citet{press} to the light-curves of the consecutive March 2006 observations. The resulting periodograms are displayed in Figure \ref{scargle_fuse}. We also plot on each periodogram a dashed line corresponding to the 99\% significance level for variability based on white noise simulations. The FUV continuum periodogram shows a significant peak at a frequency of 3.15 cycles day$^{-1}$ (P=7.6$\pm$0.4~h), and another minor peak barely exceeding the 99\% significance threshold at a frequency of 2.30 d$^{-1}$ (P=10.4$\pm1.4$~h). The \ion{O}{6} P~Cygni blue edge periodogram shows a significant peak at a frequency of 2.93 d$^{-1}$ (P=8.2$\pm$0.5~h), and a second significant peak, just slightly weaker, at a frequency of 1.55 d$^{-1}$ (P=15.5$\pm$2.5~h). This second peak might be linked to the first one since its frequency is almost 1/2 of the frequency of the main peak (the slight difference is perhaps just an effect of the limited time coverage of the data). The situation is less clear for the \ion{S}{6} P~Cygni absorption periodogram, partly due to the poor time coverage of the \ion{S}{6} light-curve for observation E1130105 (during which SiC1B and SiC2A greatly suffered from pointing losses). There are two peaks around a frequency of 3~d$^{-1}$ (8.00~h), but they just only slightly exceed the 99\% significance level.

The errors on the above periods were estimated as the HWHM of gaussian fits to the peaks in the Lomb-Scargle periodograms. Given the large widths of these peaks and the modest size of the data set from which we computed the periodograms, we will refer to the above measurements as dominant time scales and not as periods. From the results above, the variations in the FUV continuum and in the \ion{S}{6} and \ion{O}{6} P Cygni profiles are all consistent with a dominant time scale of $\sim$8~h.

\subsection{{\it XMM-Newton} spectrum \label{XMM_spec}}

Even with the modest energy resolution of the EPIC instruments, it is clear from Figure \ref{epic} that the X-ray spectrum of WR~46 is dominated by emission lines. Given the weak \ion{N}{4} and strong \ion{N}{5} lines in the optical and UV, the presence of \ion{N}{6} and \ion{N}{7} is not unexpected and it is clear that these ions account for most of the peak near 500 eV. \ion{N}{6} and \ion{N}{7} appear much stronger than the neighboring \ion{O}{7} and \ion{O}{8} lines near 600 eV, where the spectrum reaches a local minimum. The X-ray O/N ratio, although uncertain because of the poor resolution, looks roughly consistent with that seen in the optical. \citet{gosset} modeled the X-ray spectrum with a combination of three optically-thin variable-abundance equilibrium plasmas of different temperatures. We took the alternative approach of building a model from individual ions from the K-shell emission of H-like and He-like ions and L-shell emission from \ion{Fe}{17}-\ion{}{18}. As shown in Figure 6, this reproduces the observed spectrum well, giving the best-fit line intensities shown in Table \ref{linelist}. This table presents one entry per ion for the strongest line. There are many more lines in the model, but EPIC cannot resolve them. For \ion{Fe}{17} and \ion{Fe}{18}, which both contribute many lines to WR~46, the relative intensities of all the lines were fixed to that of the strongest line, giving one free parameter. Similarly, EPIC cannot resolve the He-like triplets, so there is also one free parameter only each for \ion{N}{6}, \ion{O}{7}, \ion{Ne}{9}, \ion{Mg}{11}, \ion{Si}{13}, \ion{S}{15}, \ion{Ar}{17}. Because of the resolution, the line fluxes are only intended to be a rough guide to what future instruments might find.

\subsection{{\it XMM-Newton} light-curves and period search \label{xmmlc}}

Figure \ref{x_lc} displays the simultaneous X-ray and UV light-curves obtained in the present work. We can see that the amplitude of the X-ray light-curve is around $\pm$20\%. The UV light-curve has quite an irregular shape and shows variations on the order of $\pm$5\%. We again applied the fast Lomb-Scargle algorithm \citep{press} to search for periodicities in these light-curves. The resulting periodograms are displayed in Figure \ref{scargle_xmm}, with dashed lines still corresponding to the 99\% significance level for variability.  The UV periodogram shows a significant peak at a frequency of 2.84 d$^{-1}$ (P=8.5$\pm$1.8~h), while the X-ray periodogram shows a significant peak at a frequency of 2.66 d$^{-1}$ (P=9.0$\pm$2.5~h).

From the {\it XMM-Newton} data, \citet{gosset} extracted light-curves in various energy bands. They found a significant period of 7.9~h in the very soft X-rays (0.2-0.5 keV) and no significant variability at higher energies (above 0.5~keV). Our time scale for the X-ray variability in the total {\it XMM-Newton} band is consistent with that found by \citet{gosset} for their 0.2-0.5~keV light-curve, not surprisingly since the very soft photons contribute to a large proportion of the counts in our EPIC-pn light-curve. Since the X-ray spectrum is dominated by lines (\S\ref{XMM_spec}), the variability seen in the soft X-rays is also in lines, although it is not really possible to tell whether this is due to emission or absorption and which lines in particular are variable.

Given the small number of cycles covered by the observations and the large widths of the peaks of the periodogram, we can only conclude that both the X-ray and UV light-curves vary on a dominant time scale close to 8-9~hours. These results are not incompatible with the variability time scale observed in the FUV continuum and P~Cygni absorption components. We stress however that the two data sets are separated by more than four years.

\subsection{Cross-correlation analysis \label{corr}}

Using the light-curves shown in the previous sections, we checked for possible correlations and delays between the FUV continuum and the P Cygni absorption component variations, and between the X-ray and UV variations. Such an analysis could eventually be useful to constrain a detailed model of the variability of WR~46. To test the overall similarity of the light-curves as a function of time lag, we shifted in time the simultaneous light-curves with respect to each other and calculated the correlation coefficient $r$ \citep{num_rec} using the IDL routine C\_CORRELATE.PRO. Before we did so, we rebinned the {\it XMM-Newton} X-ray and UV light-curves into common time bins. All the {\it FUSE} light-curves already had common time bins.

In Figure \ref{ccor}, the cross-correlation functions are shown for the different light-curves compared. In the first case, we shifted the X-ray light-curve by $\Delta t$ and calculated the correlation with the OM UV light-curve. In the other two cases, we shifted the \ion{O}{6} and the \ion{S}{6} light-curves by $\Delta t$ and calculate the correlation with the FUV continuum light-curve. Note that this means that delays with respect to the UV or FUV variability correspond to negative values of $\Delta t$ is this figure.

We see from the plots of Figure \ref{ccor} that consecutive positive correlation peaks are roughly spaced by the dominant variability time scale of 8 hours, and the same thing is true for consecutive negative correlation (anti-correlation) peaks. The cross-correlation analysis also tells us that the \ion{O}{6}, \ion{S}{6} and X-ray light-curves are more similar in shape to the FUV/UV light-curve when a shift is applied, suggesting that if they are related, there are possibly time delays between the variations in the continuum and the variations in the line-formation regions.

\section{Discussion \label{discuss}}

\subsection{FUV spectroscopic variability and CIRs\label{gen}}

The variability seen in excess of $v_{\infty}$ in P~Cygni profiles is usually attributed to variable amounts of rarefied gas accelerated to velocities higher than $v_{\infty}$. The shocks associated to this velocity excess can either originate from small-scale stochastic fluctuations coupled to the radiative instability intrinsic to line-driven winds \citep{owo88, gayley95}, or from large-scale quasi-periodic variability that can be induced by changes in the underlying star \citep[e.g.][]{cranmer96}.

The variability time scale of $\sim$8~h that we found in the blue edge of the \ion{O}{6} doublet (and the \ion{S}{6} doublet to some extent) in WR~46 is close to the photometric and spectroscopic periods previously reported. It is also consistent with the variability time scale observed in the FUV continuum, suggesting that the FUV spectroscopic variability is not stochastic but caused by a large-scale perturbation controlled by an underlying quasi-periodic clock.

CIRs are relatively large-scale spiral-shaped density and velocity perturbations that extend from the stellar surface to possibly several tens of stellar radii into the stellar wind \citep{mullan84}. The Discrete Absorption Components (DACs) observed in the (unsaturated) absorption part of the UV P Cygni profiles in hot stars are commonly assumed to be a manifestation of CIRs. If the short-term variability of WR~46 is associated to CIRs, we might expect to see Discrete Absorption Components (DACs) in the unsaturated \ion{S}{6} P~Cygni profile of WR~46. However, given the low signal-to-noise, sparse time coverage, and wealth of strong interstellar lines in this spectral region, it is not possible to identify in the \ion{S}{6} doublet the velocity displacement of an absorption feature which is typical of DACs (see Fig. \ref{gray}).

In addition to producing DACs, modeled CIRs can also modulate the blue edge of the absorption component of saturated P~Cygni profiles on the same time scale \citep{cranmer96}. This is in line with observations showing that the steep blue edge of saturated UV lines varies in all cases when DACs are found in other lines \citep{kaper96}. Based on the modulations observed in the blue edge of the \ion{O}{6} P~Cygni absorption component of WR~46, we suggest that the short-term variability of the star might be related to CIRs, even though a typical DAC behavior could not be clearly identified in the unsaturated \ion{S}{6} line given our limited dataset. It is worth noting the case of the WR star EZ~CMa (WR~6), for which 16 consecutive days of monitoring with {\it IUE} ($>$4 cycles) helped to link the observed UV P~Cygni profile variability to CIRs \citep{stlouis95}. For this star, the variability is mainly found in the blue edge of the P~Cygni absorption troughs.

\subsection{Variability originating from the photosphere \label{photo}}

The recent model atmosphere of WR~46 by Crowther (private communication) shows that the optical depth of the stellar continuum at 1000~\AA\ reaches a value of $\sim$1 at the very base of the wind, and similarly for the 5000~\AA\ continuum. The FUV and optical continuum photometry, which both show variability at a $\pm$5\% level, are therefore dominated by the deep layers of the stellar wind, close to the hydrostatic core. The absence of core eclipses in the FUV (\S\ref{fuv_lc}) and optical continuum light-curves \citep[e.g.][]{veena} suggests that the variability itself also originates from the very base of the wind and not from a binary companion.

The similar periods or time scales ($\sim$8 h) found here or in previous studies in the optical, UV, FUV, and X-ray photometric and/or spectroscopic variability suggest that these are all related in some way. Even though they are not in phase (albeit correlated), the fact that the amplitude of variability in the \ion{S}{6} and \ion{O}{6} FUV lines is larger when the amplitude of the FUV continuum light-curve is also larger (see Fig. \ref{gray}, \ref{FUV_lca} and \ref{FUV_lcb}) also suggests a link between the continuum variations and line variability further out in the wind. The variability of WR~46 could originate from the photosphere of the star and propagate out, modulating the absorption of X-rays or the structure of wind shocks on the same time scale as the $\sim$8-h period. 

If this is correct, we might observe delays between the continuum variations and the other manifestations of the variability. According to the CMFGEN model of WR~46, the peak of the line formation region of the \ion{O}{6} doublet is around 2.5~$R_{*}$, while the formation of the  \ion{S}{6} doublet is much closer to the photosphere. \citet{gosset} also estimated that the \ion{N}{6} and \ion{N}{7} X-ray lines, which we argued dominate the X-ray spectrum and the X-ray variability (see \S\ref{XMM_spec}), form above a few stellar radii. Given the parameters adopted in Table \ref{param}, we should expect the line variability (in particular the \ion{O}{6} and X-ray variability) to lag the continuum variability by a few hours if perturbations are advected by the wind. The fact that the cross-correlation functions for the \ion{O}{6} and X-ray light-curves (Figure \ref{ccor}) do not show an extremum when the time lag is zero suggests that these are not in phase with the continuum variability. The presence of extrema corresponding to delays of a few hours are also not inconsistent with the expected time lags. 

It is worth noting that \citet{berg96} found correlated variability in the X-ray and H$\alpha$ emission from the O4If supergiant $\zeta$ Puppis. They suggested this was evidence for periodic density variations propagating from the base of the wind to further out where X-rays are produced, an interpretation that is very similar to what we propose for WR~46.

\subsection{Lack of stability of the period}

We did not find any significant long-term changes in the {\it FUSE} spectra between 2000 and 2006 (in continuum and line strength). As a consequence, we found no evidence for a change of the mass-loss rate, unlike the change in the global mass-loss rate in the early 90's reported by \citet{veen02b}. The variability time scales of our 2006 FUV data are consistent with the dominant photometric and spectroscopic periods found by \citet{oliveira2004} in their data from 1998-1999. It is therefore possible that while the global mass-loss rate did not change, the dominant period of the system also stayed the same between 1998 and 2006. This is not incompatible with the suggestion that the long-term changes (luminosity, mass-loss rate) in WR~46 are linked to the period changes \citep{veen02b, veenc}.

Inspection of Figure \ref{periods} reveals that there is no clear trend (either increasing or decreasing) in the evolution of the period of WR~46. From two intense data sets, \citet{veena} showed that there is a decrease of the photometric period from 6.78~h to 6.54~h between 1989 and 1991, and that this period change is significant. Then, most of the periods reported for later years are longer, although it is hard to assess the reliability or precision of some of these periods, especially when no uncertainties were quoted. Interestingly, the time scales that we measured from the FUV continuum and \ion{O}{6} blue edge light-curves (see \S\ref{fuv_lc}) are significantly longer than the dominant photometric periods found by \citet{veena}. Even when considering the rather large uncertainties on our measured time scales, our values are not consistent with the periods from 1989 and 1991.

This apparent lack of stability of the period of the system is a strong argument against rotational modulation as the single origin of the short-term variability of WR~46. A well-known example where rotational modulation has been linked to the short-term variability is the case of EZ~CMa (WR~6) \citep{stlouis95, morel98, flores}. This star always displays the same period of 3.766 days, even though its variability is epoch-dependent. If rotational modulation is responsible for the variability, the period should indeed always be the same or should change very little and gradually over time. We can also argue that rotation of a single WR star is not a likely cause of short-term variability with a period of a few hours because it would generally require a rotational velocity that is high relative to the break-up velocity of the star. For WR~46, assuming a rotation period of 8~h and the parameters of Table \ref{param}, we find that the star would rotate at about 40\% of its critical rotational velocity at the equator, with a velocity of $\approx 440$~km~s$^{-1}$. This is not impossible per se but still rather high compared to the first estimates of rotation rates of WR stars by \citet{chene08,chene2010} who found rotational velocities typically of a few tens of km~s$^{-1}$.

The lack of stability of the period is also difficult to explain in terms of binarity alone. After considering conservation laws (spin and orbital angular momentum, energy) and various exchange mechanisms, \citet{veenc} did not exclude that the observed period change from 1989 to 1991 represents an orbital change due to the spiral-in of a companion, but they considered such a scenario highly unlikely. Between 1989 and 1991, the decrease in the photometric period is 14 minutes \citep{veena}, which is extremely large for any binary system \citep[see][]{veenc}. This spiral-in scenario seems even more unlikely when we consider the subsequent measurements of longer (and possibly multiple) periods. Again, the time scales that we measured from our 2006 {\it FUSE} observations are significantly longer than the dominant photometric periods found in 1989 and 1991 by \citet{veena}.

\subsection{X-rays and the Hatchett-McCray effect \label{hm}}

If we compare the X-ray properties of WR~46 with those of the three WR+compact systems known, it seems very unlikely that WR~46 belongs to the same class. The most famous WR+compact system is probably the strong galactic X-ray source Cyg-X3 ($L_{\rm X} \approx 10^{38}$~erg~s$^{-1}$, \citealt[e.g.][, and references therein]{schmu96}). It has an orbital period of 4.8 h \citep{parsignault72} and is thought to consist of a WR star \citep{vankerk92} possibly accompanied by a black hole \citep{schmu96, han2000}, although this has been contested \citep[e.g.][]{mitra}.  A first extragalactic WR+compact candidate, IC~10~X-1, was detected in the starburst galaxy IC~10 \citep{bauer2004, wang05}. It also has a high X-ray luminosity ($L_{\rm X} {\rm (0.2-10~keV)}\sim 1.2 \times 10^{38}$ erg s$^{-1}$), and its period is close to 35~h \citep{prest07}. \citet{silver} confirmed that this system is a WR + black hole binary. Another extragalactic candidate, NGC~300~X-1, was identified by \citet{carpa2007a, carpa2007b} and then confirmed as a WR + black hole binary by \citet{crowther2010}. Its observed luminosity in the 0.2-10~keV band is $\sim 2 \times 10^{38}$~erg~s$^{-1}$, and it was found to have a period close to 32~h. As well as the orders of magnitude higher luminosities and much harder spectra, those accreting X-ray sources show a much smoother continuum spectrum compared to the line-dominated X-ray spectrum of WR~46 (Figure \ref{epic}). Again, we stress that the X-ray spectrum and luminosity of WR~46 are typical of single WN stars.\\

Estimating the total X-ray luminosity produced by Bondi-Hoyle accretion of a stellar wind onto a degenerate object \citep[e.g.][]{stevens88} for the parameters of WR~46 (assuming a circular orbit with P=8~h), we obtain $L_{\rm X}\sim 10^{37}$~ergs~s$^{-1}$. This expected X-ray luminosity is several orders of magnitude higher than what is observed for WR~46, and this is very unlikely caused by wind attenuation of the X-rays. For example, the predicted X-ray accretion luminosity of EZ~CMa is $\sim 10^{36}$~erg~s$^{-1}$, about 3 orders of magnitude higher than what is observed, an argument that has been used against an accreting neutron star companion scenario \citep{skinner97}. For this star, \citet{stevens88} performed numerical calculations accounting for wind attenuation for a range of system parameters and they predicted values of $L_{\rm X} \sim 10^{35}-10^{36}$~erg~s$^{-1}$, still much larger than the observed value. WR~46 is a weak-lined WN star and its wind is less dense than that of EZ~CMa, so the effect of wind attenuation should not be stronger than in EZ~CMa. Thus, it is highly unlikely that the wind of WR~46 hides a very luminous X-ray companion, especially one that would imply a $>$4 orders of magnitude difference between the intrinsic and absorbed X-ray luminosity.

Various suggestions like centrifugal inhibition of accretion \citep{Illarionov75, stella86} have been discussed as potential mechanisms that could explain the lack of X-rays from a compact companion. For WR~46, \citet{veenc} also discussed other possible types of companion like a low-mass main sequence star, which would however imply an unexpected initial mass ratio \citep[e.g.][]{garmany}. Some of these more unusual types of companions might not be very luminous in X-rays. However, in such cases, even though the observed level of X-rays could possibly be reconciled with the presence of a companion, its impact on the stellar wind would remain minimal as we illustrate below. 

We have already reviewed many problems associated with the interpretation of the variability of WR~46 in terms of binarity alone. Nevertheless, assuming that more than one cyclical mechanism could be at play in this star, it is worth checking if a companion could be responsible for the variability observed in the P~Cygni profiles. An X-ray source immersed in the stellar wind of its companion star can ionize a surrounding region and lead to observable variations. In particular, such a system could show variability with orbital phase in the P~Cygni profile of ultraviolet resonance lines, an effect known as the Hatchett-McCray effect \citep{hm77}, which has been observed in several high-mass X-ray binaries (HMXRBs) \citep[e.g.][]{vanloon}.

If an X-ray source was present in the wind of WR~46, then there would be a region around this source in which atoms are ionized to higher stages. In this region, the fraction of \ion{O}{6} or \ion{S}{6} with respect to other ionization stages would change. For a given transition, an increased number of scatterers in the ionization region would lead to an enhanced P~Cygni absorption and/or emission, depending on where the ionization region is with respect to the WR star and the observer. Similarly, if the number of scatterers decreases in the region surrounding the X-ray source, the P~Cygni absorption and/or emission weakens. Note that it is possible for the variations in the P~Cygni emission component to remain unnoticed while variations are observed in the absorption component, because the region ionized by the X-ray source represents a smaller fraction of the volume contributing to the emission component compared to the fraction of the volume contributing to the absorption component.

Several factors however appear to go against an interpretation of the variability of the P~Cygni profiles of WR~46 in terms of the Hatchett-McCray effect. First of all, the X-ray luminosity of WR~46 ($\sim$10$^{32}$ erg~s$^{-1}$) is very low compared to the typical X-ray luminosity of known HMXRBs showing the Hatchett-McCray effect ($\sim$10$^{35}$-10$^{38}$ erg~s$^{-1}$) \citep[e.g.][]{vanloon}. Also, since the size of the ionized region for a given species and ionization stage scales linearly with the X-ray luminosity of the companion and inversely with the density of the wind \citep{hm77}, we expect this region to be very small for WR~46 which has a relatively low X-ray luminosity and a denser wind than typical O stars. We can estimate the size of such an ionized region by considering the quantity \citep{hm77}

\begin{equation}
\xi(r, r_{\rm X}) = \frac{ L_{\rm X}}{n(r) \ r_{\rm X}^2} = \frac{4 \pi \ L_{\rm X} \ \bar{m}}{\dot{M}} \ v(r) \left( \frac{r}{r_{\rm X}} \right)^2,
\label{xi}
\end{equation}

\noindent{where $n(r)$ is the local number density of the gas, $r_{\rm X}$ is the distance from the X-ray source, $\bar{m}$ is the average mass per ion (we assume here a pure helium atmosphere), $L_{\rm X}$ is the intrinsic (unabsorbed) X-ray luminosity of the companion, $v(r)$ is the velocity of the WR wind, and $\dot{M}$ is the WR mass-loss rate. In the limit $\xi \rightarrow 0$, the ionization balance of the material is unaffected by the presence of the X-ray companion and thus mainly governed by the radiation field of the WR star. According to the ionization models of \citet{kallman}, significant changes in ionization fraction generally only occur for values of $\log{\xi} > 1.6$. The variations of $\log{\xi (r, r_{\rm X})}$ in the orbital plane for the case of WR~46 are shown in Figure \ref{lx_ion} for an assumed intrinsic X-ray luminosity of $10^{33}$, $10^{34}$, or $10^{35}$~erg~s$^{-1}$.
For this calculation, we assumed an orbital period of 8~h and a circular orbit, a $\beta=1$ velocity law, and we used the values of Table \ref{param}. The corresponding distance of a compact companion is 5.3 $R_{\sun}$, or 1.83 $R_{*}$.

The Hatchett-McCray effect is obviously more difficult to observe if the ionized region is small. If it is small enough, variability should be limited to the absorption components of P~Cygni profiles and to the fraction of the orbit during which the companion is between the star and the observer. We can see from Figure \ref{lx_ion} that even in the rather unlikely case where the unabsorbed X-ray luminosity of the companion is $10^{35}$~erg~s$^{-1}$ (see discussion above), the region within which $\log{\xi} \geq 1.6$ is relatively small, and it is unlikely to perturb the column in front of the star for most of the orbit. Note also that equation \ref{xi} is valid for an optically thin gas, and that in reality the mean path length of the X-ray photons should be considerably shorter. The X-ray ionized zone is thus likely even less extended than what is shown in Figure \ref{lx_ion}. The fact that we see spectral variations during a major fraction of the $\sim$8-hour cycle appears in contradiction with the Hatchett-McCray effect of a small ionized region. The ionized region would also have to be much larger than predicted here in order to encompass the region of the wind between -0.6~$v_{\infty}$ and -$v_{\infty}$, where variability is observed in the \ion{S}{6} P~Cygni absorption.

\subsection{Nonradial pulsations}

The variability observed in this work in the FUV continuum and in the UV, as well as the optical continuum variability seen by other authors, would find a natural explanation in the context of pulsations. It could either be caused directly by brightness variations at the stellar surface, or by velocity/density perturbations at the stellar surface deforming the continuum forming layer. As discussed in \S\ref{photo}, this could then induce variability further out in the wind.

The scenario of non-radial pulsations is also appealing to explain the multiple photometric periods present at a given epoch \citep{veena, oliveira2004}. Moreover, the different periods measured and the fact that the variations are not strictly periodic might easily be explained by the presence of multiple pulsation periods. The interference patterns between different modes could result in a complex light-curve, varying both in shape and amplitude. It is also conceivable that a time-limited data set would allow to identify only one of the periods, and that another time-limited data set would show another of the periods. 

Many studies have explored and discussed the possibility of pulsations in Wolf-Rayet stars \citep[e.g.][]{noels, glatzel1993, lefevre2005, townsend06, dorfi, glatzel08}. For WR~46, \citet{veenc} proposed a simple geometrical model in which non-radial pulsations would be able to cause radial-velocity shifts of the wind emission lines. They suggested that the lowest order sectoral mode $l$=1 and $|m|$=1, which results in a single bright hemisphere travelling along the equator, could produce a distorsion of the wind of WR~46 (see their Fig. 5) and lead to a double-wave photometric period and a single-wave radial-velocity period, as observed.  The bright hemisphere could distort the wind by causing an asymmetric density flow, with the ``one-armed''  wind appearing to rotate as the pulsational pattern travels over the stellar surface. In such a scenario, non-radial pulsations could generate CIRs in the wind.

\section{Conclusion \label{conc}}

From our time-resolved {\it FUSE} observations of WR~46, we detected ``blue edge'' variability in the \ion{O}{6} doublet P~Cygni absorption and line-profile variations in the \ion{S}{6} doublet P~Cygni absorption. Both variations occur on a time scale of about 8~h, consistent with the time scale found in the FUV continuum light-curve. A similar time scale was also recovered in UV and X-ray light-curves from archival {\it XMM-Newton} observations.

We noted that the X-ray spectrum and luminosity of WR~46 are typical of a single WN star. If a companion with a lower X-ray luminosity like a rapidly rotating neutron star was present, its effect on the stellar wind would be minimal, and its limited ionization effect (the Hatchett-McCray effect) could not explain the variability observed in the FUV P~Cygni profiles.

We also pointed out that the fact that the observed period is not stable (changing with time with no particular trend) is a strong argument against rotational modulation or binarity alone as the origin of the short-term variability.

Circumstantial evidence suggests that the variability of WR~46 is rooted in photospheric perturbations, and that CIR-like structures might induce the observed P~Cygni line profile and X-ray line variability. We proposed that non-radial pulsations might be the source of these perturbations, although an improved theoretical understanding of non-radial pulsations in WR stars and how they propagate in the stellar wind is required to really investigate this possibility further.

Meanwhile, we suggest that additional intensive and long-term monitoring campaigns of WR~46 would certainly prove useful, for example to follow the evolution of the period or the recurrence of multiple periods. If possible, continuous photometry for several days or weeks would be a perfect way to test the scenario of non-radial pulsations with multiple periods.

\acknowledgments
We are thankful to Paul Crowther for providing results from his recent wind model of WR~46. We also thank the anonymous referee for comments which helped to improve the paper. V.~H.-B. acknowledges support from NSERC (Canada) for a postgraduate scholarship. N.~S.-L. is also grateful to NSERC for financial assistance.


%% in parentheses, after the keyword, but they will not be verified.

%{\it Facilities:} \facility{}}.



\appendix

\clearpage

\begin{figure}
\epsscale{.7}
\plotone{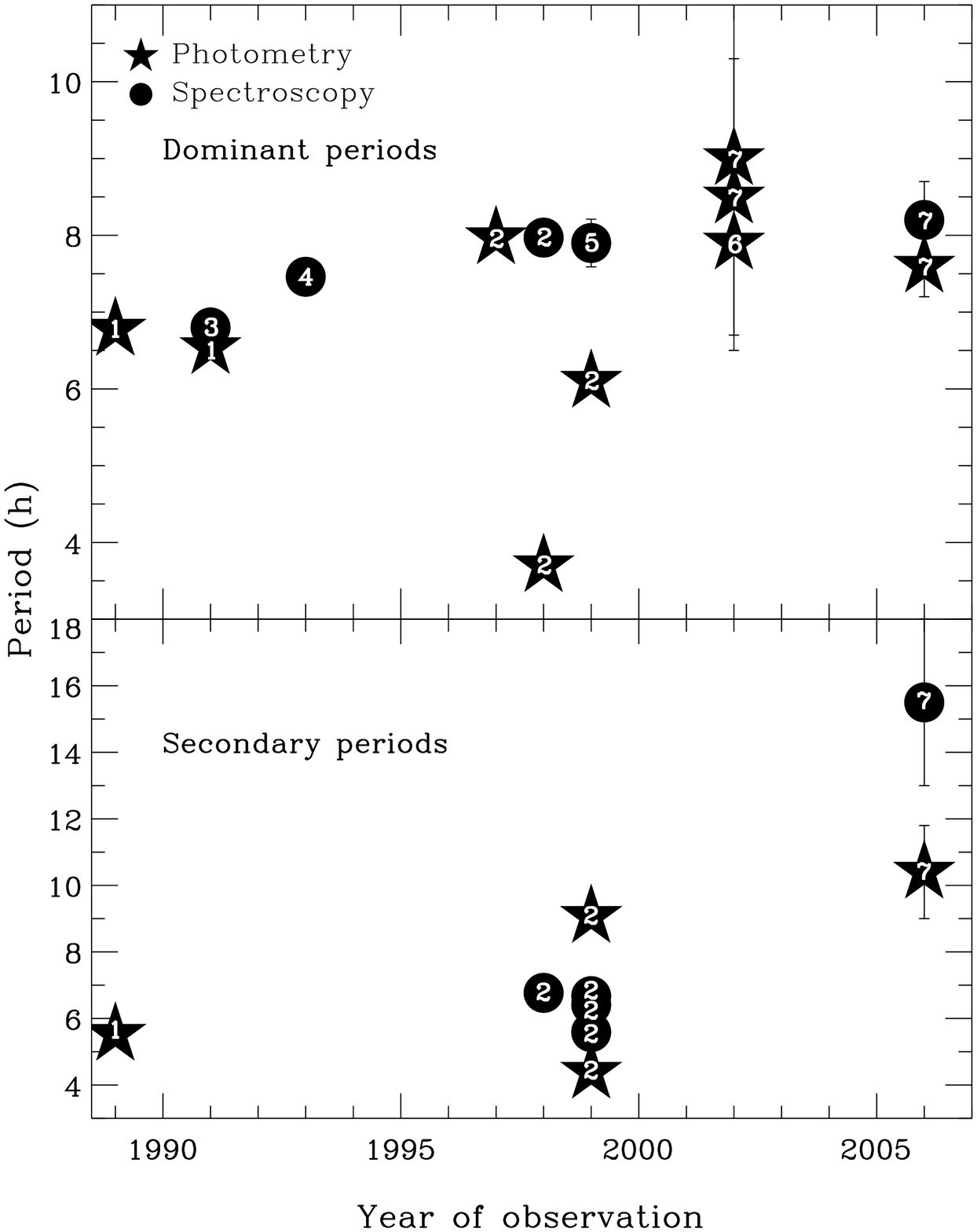}
\caption{Periods reported for WR~46 plotted versus year of observation. The photometric periods are shown with stars and the spectroscopic periods with circles. The upper panel shows the dominant periods for a given data set and the bottom panel shows periods with weaker amplitudes. The references below are identified on each measurement by their number. Error bars are shown when available. {\it References $-$} (1) \citet{veena} , (2) \citet{oliveira2004}, (3) \citet{veen95}, (4) \citet{niemela95}, (5) \citet{marchenko2000}, (6) \citet{gosset}, (7) This work.
   \label{periods}}

\end{figure}

\begin{figure}
\epsscale{.85}
\plotone{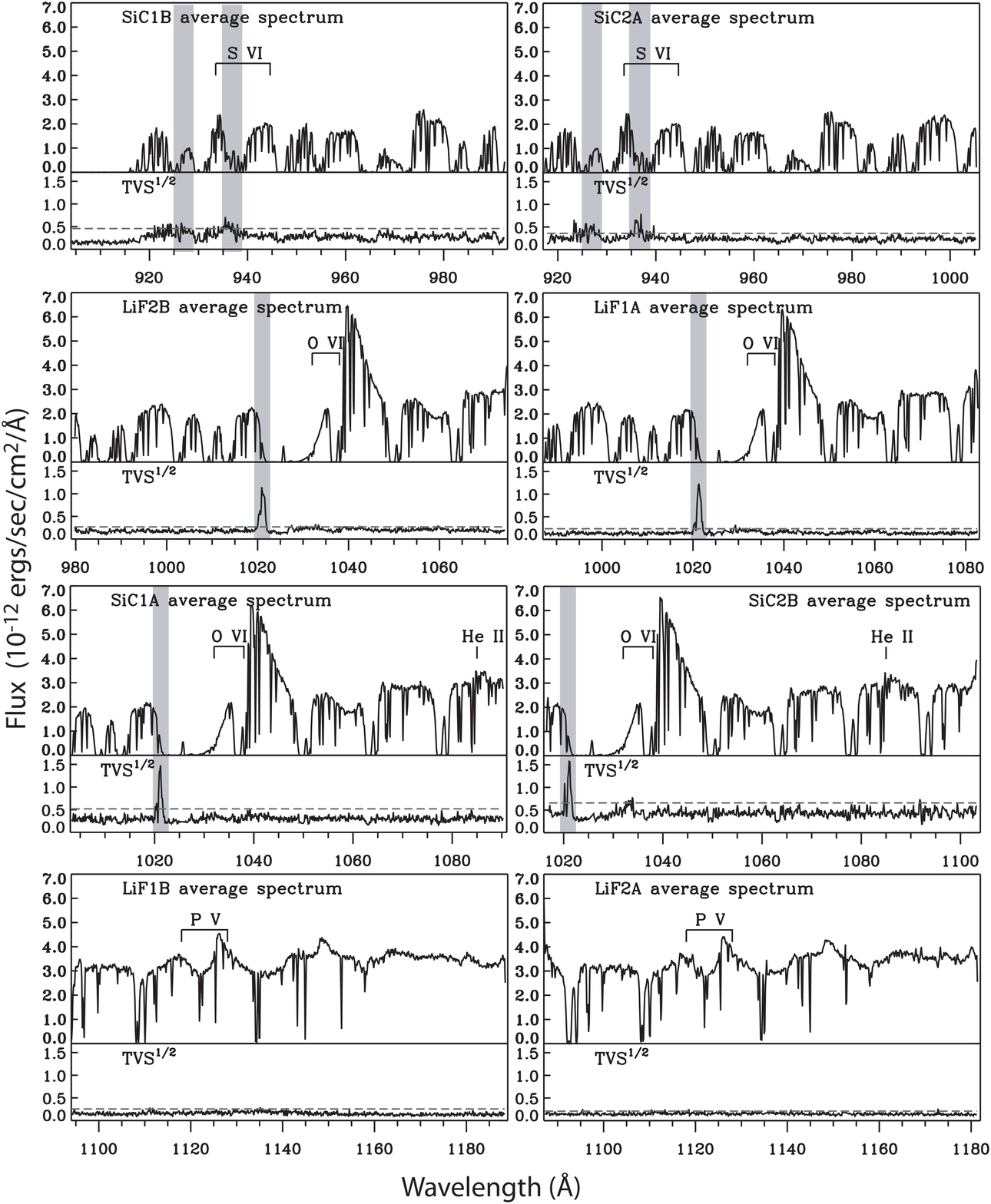}
\caption{{\it Top panels:} Time-average spectrum for each {\it FUSE} channel. {\it Bottom panels:} Square-root of the temporal variance spectrum (TVS$^{1/2}$) of our time series of spectra. The dashed line indicates the 99\% confidence level for variability. The wavelength intervals with significant spectral variability and for which light-curves were computed are highlighted in gray.}
\label{tvs}
\end{figure}

\begin{figure}
\epsscale{.88}
\plotone{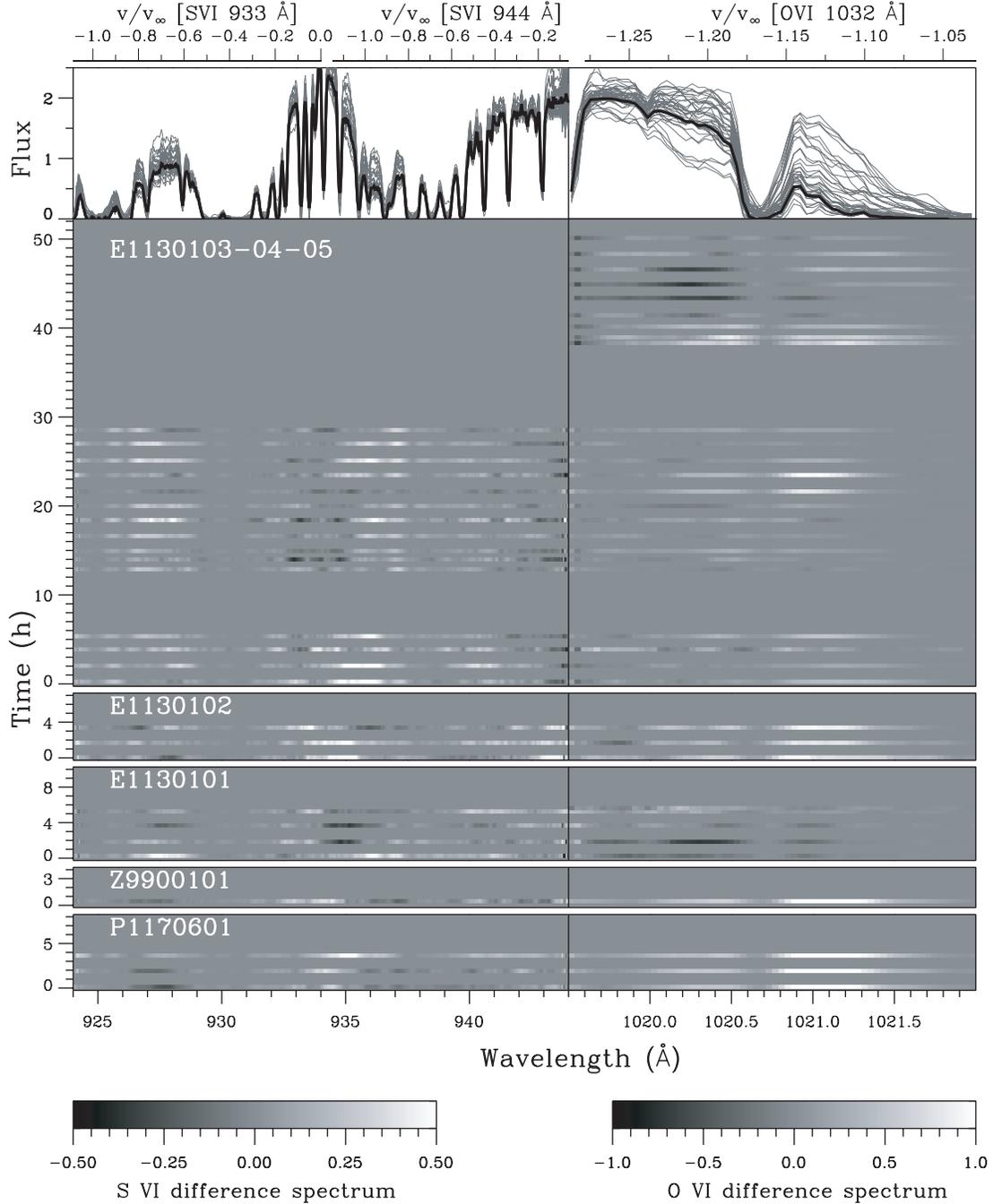}
\caption{Gray-scale plot of the differences between individual spectra and the time-averaged spectrum ({\it top}, thick line) for the \ion{S}{6} $\lambda\lambda$933, 944 P Cygni absorption trough ({\it left}) and the \ion{O}{6} $\lambda\lambda$1032, 1038 P Cygni absorption blue edge ({\it right}). Time is given in hours since the beginning of each observation. The velocity scales above the top panel are given with respect to the center of the \ion{S}{6} $\lambda$933,  \ion{S}{6} $\lambda$944, and \ion{O}{6} $\lambda$1032 lines, using the adopted $v_{\infty}=2775$ km s$^{-1}$. The narrow absorption features are of interstellar origin. Smoothing factors of 15 and 5 were applied to the \ion{S}{6} and \ion{O}{6} difference spectra respectively. The real signal-to-noise of the difference spectra is therefore lower than seen here. Fluxes are in units of $10^{-12}$~ergs~s$^{-1}$~cm$^{-2}$~\AA$^{-1}$. } 
\label{gray}
\end{figure}

\begin{figure}
\epsscale{1.0}
\plotone{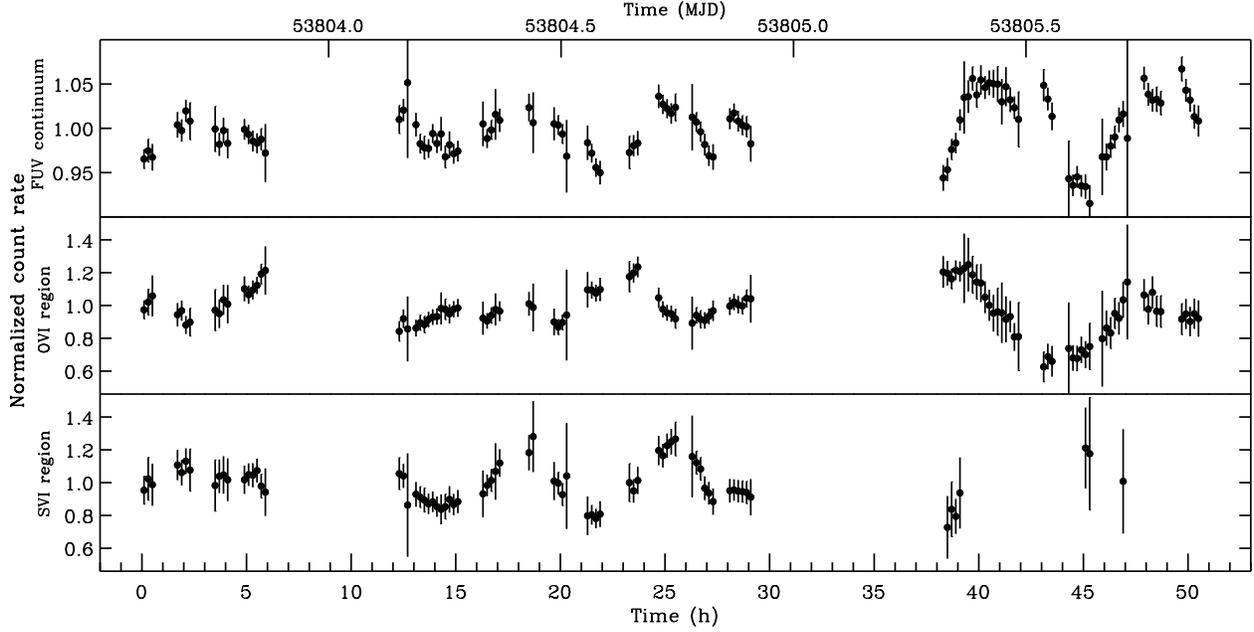}
\caption{Light curves for the consecutive March 2006 observations in spectral windows corresponding to the FUV continuum ({\it top panels}) and the variable regions of the \ion{O}{6} and \ion{S}{6} P Cygni profiles ({\it bottom two panels}).}
\label{FUV_lca}
\end{figure}

\begin{figure}
\epsscale{.55}
\plotone{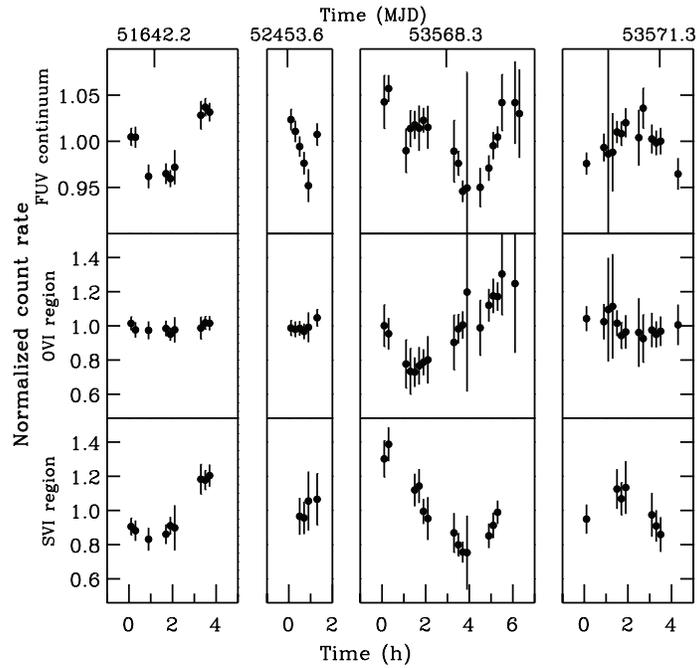}
\caption{Same as Figure \ref{FUV_lca} but for the non-consecutive observations P1170601, Z9900101, E1130101, and E1130102 respectively from left to right.}
\label{FUV_lcb}
\end{figure}

\begin{figure}
\epsscale{.8}
\plotone{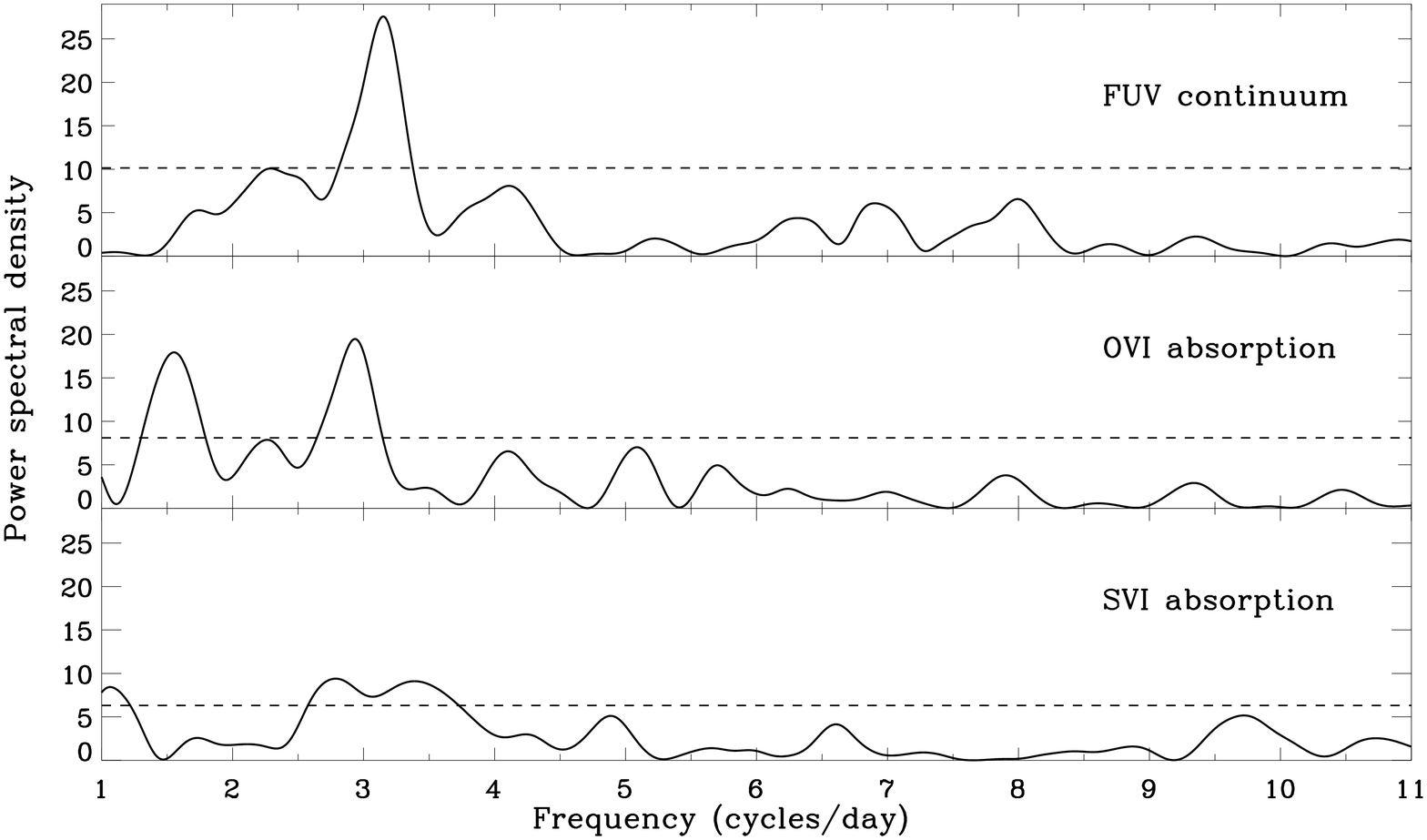}
\caption{Lomb-Scargle periodograms for the combined light-curves of March 2006 {\it FUSE} observations, in spectral windows corresponding to the FUV continuum and the variable absorption component of the \ion{O}{6} and \ion{S}{6} P Cygni profiles. The dashed lines indicate the 99\% significance level.} 
\label{scargle_fuse}
\end{figure}

\begin{figure}
\epsscale{.75}
\plotone{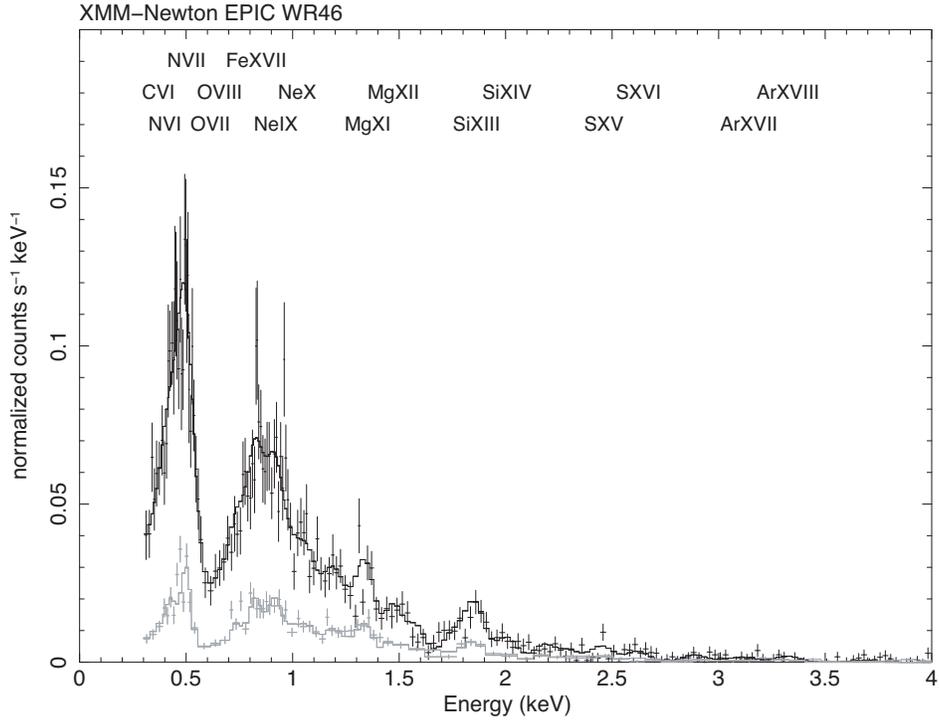}
\caption{The XMM-Newton X-ray spectrum of WR46 with EPIC-pn in black and EPIC-MOS1 and EPIC-MOS2 added together in gray. Data are shown with error bars while the continuous lines show a pure emission-line model from H-like and He-like ions of C, N, O, Ne, Mg, Si, S and Ar along with L-shell emission from FeXVII and FeXVIII.}
\label{epic}
\end{figure}

\begin{figure}
\epsscale{.8}
\plotone{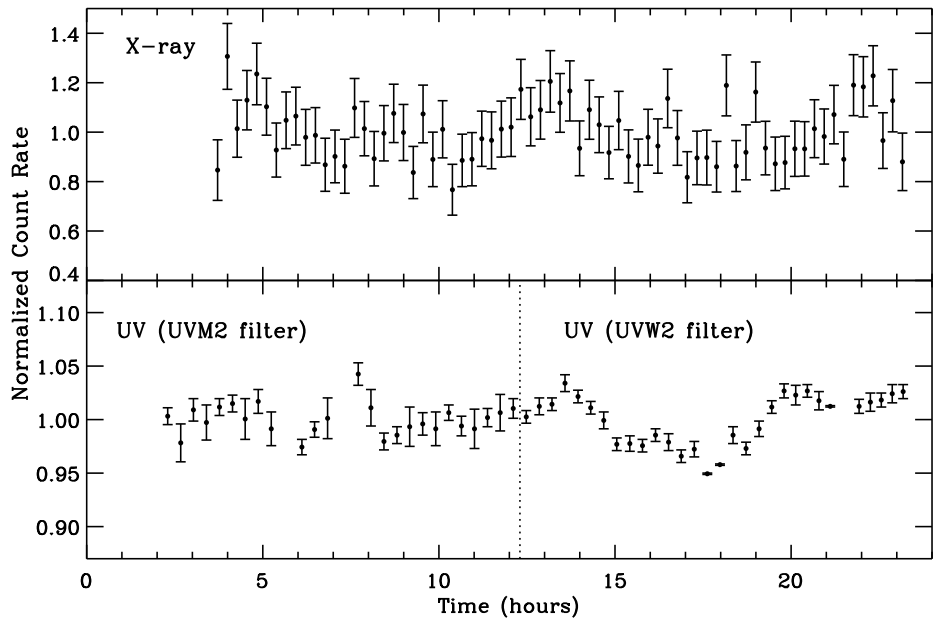}
\caption{{\it Top panel}: Broadband (0.3-10 keV) XMM-EPIC X-ray light-curve of WR~46. {\it Bottom panel}: Simultaneous XMM-OM UV light-curve using the UVM2 filter for the first half of the observation and the UVW2 for the second half. Note that before normalizing the UV light-curve, the measured rate in the UVM2 filter was converted to its equivalent in the UVW2 filter using the method described in \S\ref{xmmdat}.}
\label{x_lc}
\end{figure}

\begin{figure}
\epsscale{.75}
\plotone{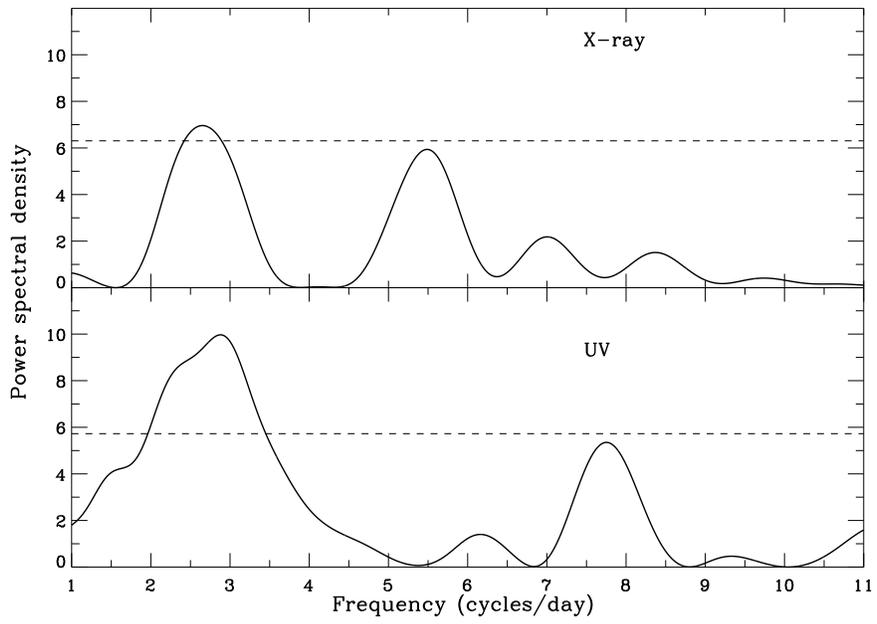}
\caption{Lomb-Scargle periodograms for the XMM X-ray and UV light-curves of WR~46. The dashed lines indicate the 99\% significance level.} 
\label{scargle_xmm}
\end{figure}

\begin{figure}
\epsscale{.5}
\plotone{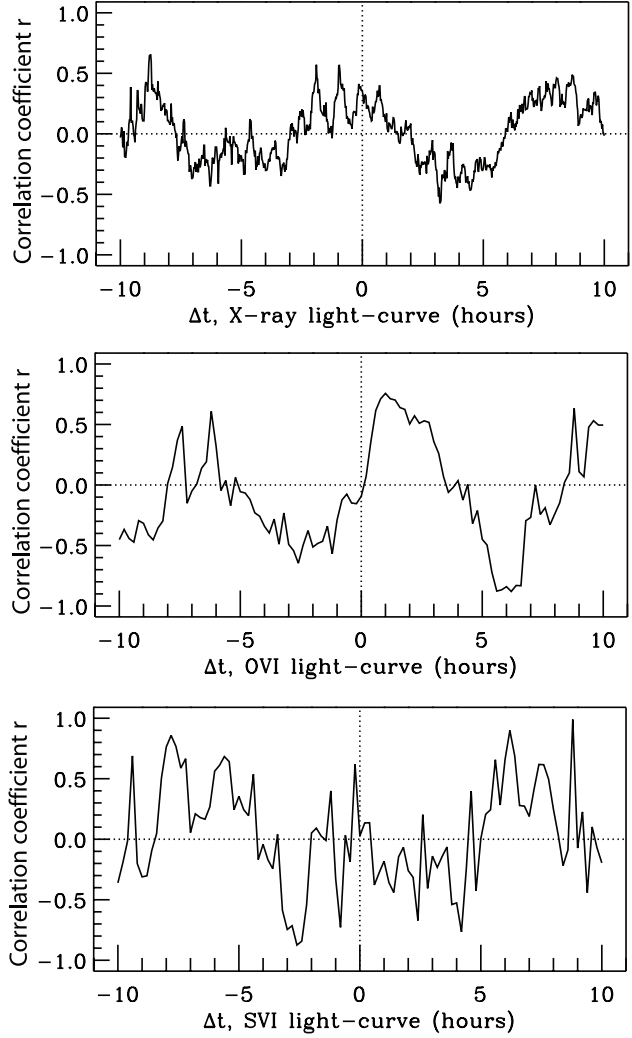}
\caption{Linear correlation coefficient $r$ as a function of the time translation applied to the light-curves. The top panel shows the correlation of the shifted XMM X-ray light-curve with the XMM UV light-curve. The middle and bottom panels show respectively the correlation of the shifted \ion{O}{6} and \ion{S}{6} {\it FUSE} light-curves with the {\it FUSE} FUV continuum light-curve.} 
\label{ccor}
\end{figure}

 \begin{figure}
\epsscale{.8}
\plotone{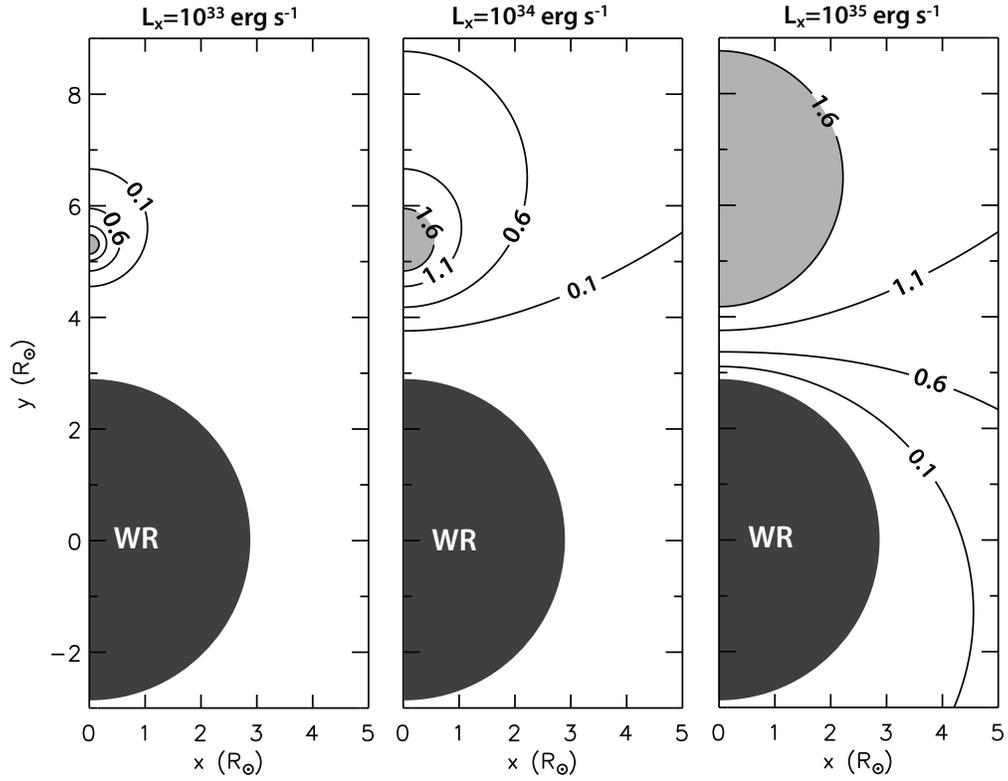}
\caption{Contours of constant ionization parameter $\log{\xi}$ (in erg~cm~$s^{-1}$) for an ionizing companion located at (0,~5.3) and having an intrinsic (unabsorbed) X-ray luminosity of $10^{33}$, $10^{34}$, or $10^{35}$~erg~s$^{-1}$ (from left to right). The Wolf-Rayet star is placed at the origin of the plots. The region within which $\log{\xi} \geq 1.6$ is indicated in light gray.}
\label{lx_ion}
\end{figure}

\clearpage

\begin{deluxetable}{cc}
\tabletypesize{\footnotesize}
\tablecaption{Adopted stellar and wind parameters of WR~46 \label{param}}
\tablewidth{0pt}
\tablehead{
\colhead{Parameter} & \colhead{Value}}

\startdata
$R_{*}$ & 2.9 $R_{\sun}$\\
$T_{*}$ & 90,210 K\\
$M$ & 18 $M_{\sun}$\\
$L$ & 0.5$\times$10$^6$ $L_{\sun}$\\
$v_{\infty}$ & 2775 km s$^{-1}$\\
$\dot{M}$ & 4$\times$10$^{-6}$ $M_{\sun}$ yr$^{-1}$\\
$d$ & 4.0 kpc \\
$L_{\rm X}$(0.2$-$10.0 keV) &$7.7\times10^{32}$~erg~s$^{-1}$\\
\enddata
\end{deluxetable}

\begin{deluxetable}{llccccc}
\tabletypesize{\footnotesize}
\tablecaption{Log of {\it FUSE} observations \label{log}}
\tablewidth{0pt}
\tablehead{
\colhead{Data set$^{*}$} & \colhead{Start time (UT)}& \colhead{End time (UT)} & \colhead{Aperture} & \colhead{MODE} &
\colhead{Number of exposures} & \colhead{t$_{\rm exp}$(s)}
}
\startdata
P1170601&	2000-04-08 03:36:26 & 2000-04-08 07:25:06 &	LWRS	&	TTAG&	4	&	4891 \\
Z9900101&	2002-06-28 14:31:54 & 2002-06-28 15:56:39 &	LWRS	&	TTAG&	2	&	3946\\
E1130101&	2005-07-17 04:09:40 & 2005-07-17 10:26:59 &	LWRS	&	TTAG&	9	&	10210 \\
E1130102&	2005-07-20 03:46:53 & 2005-07-20 08:12:58  &	LWRS	&	TTAG&	6	&	6199 \\
E1130103&	2006-03-09 14:28:48 & 2006-03-09 20:23:59  &	LWRS	&	TTAG&	4	&	9955	 \\
E1130104&	 2006-03-10 02:46:20 & 2006-03-10 19:38:25  &	LWRS	&	TTAG&	11	&	29682 \\
E1130105&	 2006-03-11 03:44:02 & 2006-03-11 17:06:18  &	LWRS	&	TTAG&	9	&	28213 \\
\enddata
\tablenotetext{* }{ The data are archived in the Multimission Archive of the Space Telescope Science Institute (MAST).}
\end{deluxetable}

\begin{deluxetable}{lcc}
\tabletypesize{\footnotesize}
\tablecaption{List of X-ray lines and measured fluxes \label{linelist}}
\tablewidth{0pt}
\tablehead{
\colhead{Ion} & \colhead{Wavelength (\AA)} & \colhead{Flux (cm$^{-2}$ s$^{-1}$)}}

\startdata
\ion{C}{6} & 33.7342 & 1.8 $\pm$ 1.0 $\times$ 10$^{-4 }$\\
\ion{N}{6} & 28.7871 & 1.8 $\pm$ 0.5 $\times$ 10$^{-4 }$\\
\ion{N}{7} & 24.7793 & 7.2 $\pm$ 0.4 $\times$ 10$^{-5 }$\\
\ion{O}{7} & 21.6020 & 2.4 $\pm$ 0.4 $\times$ 10$^{-5 }$\\
\ion{O}{8} & 18.9671 & 5.6 $\pm$ 1.0 $\times$ 10$^{-6 }$\\
\ion{Fe}{17} & 15.0150 & 6.9 $\pm$ 1.0 $\times$ 10$^{-6 }$\\
\ion{Fe}{18} & 14.2080 & 2.4 $\pm$ 0.6 $\times$ 10$^{-6 }$\\
\ion{Ne}{9} & 13.4471 & 1.2 $\pm$ 0.1 $\times$ 10$^{-5 }$\\
\ion{Ne}{10} & 12.1321 & 2.7 $\pm$ 0.4 $\times$ 10$^{-6 }$\\
\ion{Mg}{11} &     9.1688  &  3.3 $\pm$ 0.1 $\times$ 10$^{-6 }$\\
\ion{Mg}{12} &    8.4192  &  1.2 $\pm$ 0.1 $\times$ 10$^{-6 }$\\
\ion{Si}{13} &   6.6480  &  2.4 $\pm$ 0.1 $\times$ 10$^{-6 }$\\
\ion{Si}{14} &     6.1804  &  9.6 $\pm$ 2.1 $\times$ 10$^{-7 }$\\
\ion{S}{15} &      5.0387  &  3.4 $\pm$ 1.2 $\times$ 10$^{-7 }$\\
\ion{S}{16} &     4.7274  &   9.3 $\pm$ 5.8 $\times$ 10$^{-7 }$\\
\ion{Ar}{17} &   3.9488  &  2.9 $\pm$ 1.3 $\times$ 10$^{-7 }$\\
\ion{Ar}{18} &   3.7311  &  3.6 $\pm$ 1.0 $\times$ 10$^{-7 }$\\
\enddata
\end{deluxetable}

\end{document}